\documentclass[preprintnumbers,a4paper,twocolumn,floats,aps,nofootinbib]{revtex5}
\pdfoutput=1
\usepackage{graphicx}
\usepackage{amsmath,amssymb,amsfonts}
\usepackage{hyperref}
\usepackage{bbm}
\usepackage{xspace}
\newcommand{\eVdist}{\kern-0.06em}

\newcommand{\gev}{\:\text{Ge\eVdist V}}
\newcommand{\tev}{\:\text{Te\eVdist V}}

\def \beq{\begin{equation}}
\def \eeq{\end{equation}}
\def \bea{\begin{eqnarray}}
\def \eea{\end{eqnarray}}

\newcommand{\Z}[1]{\ensuremath{\mathbbm{Z}_{#1}}} 

\newcommand{\MO}{{\tt MicrOmegas}\xspace}

 \newcommand{\SARAH}{{\tt SARAH}\xspace}
 \newcommand{\SPheno}{{\tt SPheno}\xspace}

\begin{document}
\title{Dilaton domination in the MSSM and its singlet extensions}

\preprint{ZMP-HH/14-6, CERN-PH-TH/2014-023}

\author{Jan Louis$^{a,b,c}$}
\author{Kai Schmidt-Hoberg$^d$}
\email{kai.schmidt-hoberg@cern.ch}
\author{Lucila Zarate$^a$}

\affiliation{$^a$Fachbereich Physik der Universit\"at Hamburg, Luruper Chaussee 149,
D-22761 Hamburg, Germany}

\affiliation{$^b$ Zentrum f\"ur Mathematische Physik,
Universit\"at Hamburg,
Bundesstrasse 55, D-20146 Hamburg, Germany}

\affiliation{$^c$ Albert Einstein Minerva Center, Weizmann Institute of Science,
Rehovot 76100, Israel}
\affiliation{$^d$Theory Division, CERN, 1211 Geneva 23, Switzerland}

\begin{abstract} 
We analyse the current status of the dilaton domination scenario in the MSSM and its singlet extensions, taking into account
the measured value of the Higgs mass, the relic abundance of dark matter and constraints from SUSY searches at the LHC.
We find that in the case of the MSSM the requirement of a dark matter relic abundance in accord with observation severely restricts
the allowed parameter space, implying an upper bound on the superpartner masses which makes it fully testable at the LHC-14. 
In singlet extensions with a large singlet-MSSM coupling $\lambda$ as favoured by naturalness arguments the coloured sparticles should 
again be within the reach of the LHC-14, while for small $\lambda$
it is possible to decouple the MSSM and singlet sectors, achieving the correct dark matter abundance with a singlino LSP while
allowing for a heavy MSSM spectrum.
\end{abstract}

\maketitle

%



\section{Introduction}

With the discovery of the Higgs boson and no evidence for supersymmetry
there is a renewed interest in non-minimal versions 
of the supersymmetric Standard Model. In particular in singlet extensions such as the NMSSM 
an additional
neutral chiral multiplet $S$ with specific couplings to the Higgs sector 
is added to the field content of the minimal supersymmetric Standard
Model (MSSM) (see e.g.\ \cite{Ellwanger:2009dp,Maniatis:2009re} for reviews on the subject).
Originally this model was suggested as a dynamical mechanism 
to generate the supersymmetric $\mu$-term with its size being naturally
of order ${\cal O}(M_Z)$. However, due to the mixing of the scalar components
of $S$ with the Higgs sector, a Higgs mass $m_h \simeq 125$ GeV is obtained with more ease. 
This is due to an additional tree-level $F$-term contribution to the Higgs mass, which also helps alleviating the notorious fine tuning problem
of the  MSSM. This has been observed in the $\mathbbm{Z}_3$ symmetric NMSSM~\cite{Dermisek:2007yt} 
and particularly in its generalised version (GNMSSM)~\cite{Ross:2011xv,Ross:2012nr,Kaminska:2013mya,Kaminska:2014wia}, see also~\cite{Delgado:2010uj,Delgado:2010cw}.

It is of obvious interest to embed the NMSSM into string backgrounds.
Since scalar fields are abundant in string theory, singlet extensions
appear to be a promising low energy limit. 
The corresponding string model building has as usual two aspects. On the one hand, there is the need 
to construct explicit
string backgrounds with the additional chiral multiplet $S$
 and its specific  couplings.\footnote{See, for example, 
Refs.~\cite{Lebedev:2009ag,Cvetic:2010dz}.}
On the other hand, the mechanism for supersymmetry breaking 
determines the soft terms and thus the specific low energy particle spectrum.

In this paper we focus on the second aspect in that we do not attempt to construct a string background but instead study the effects of string motivated
soft terms. Specifically we concentrate on the situation
where the supersymmetry breaking occurs in the dilaton direction of the
heterotic string \cite{Kaplunovsky:1993rd}. This 
leads to a rather constrained set of (universal) soft terms 
-- termed ``dilaton dominated soft terms'' --
which were analysed within the MSSM in refs.\ \cite{Barbieri:1993jk,Brignole:1993dj,Louis:1994ht,Kraniotis:1995ke,Khalil:1995sz,Casas:1996wj,Casas:1996zi}.

It has been argued that dilaton domination in the MSSM suffers from the existence of charge and colour breaking vacua which are deeper than
the local electroweak vacuum~\cite{Casas:1996wj}. Subsequently, however, it has been realised that, while these deeper vacua exist, 
the lifetime of the electroweak vacuum is almost always considerably longer than the age of the Universe \cite{Riotto:1995am,Evans:2008zx}(see e.g.\ Fig.~1 of \cite{Evans:2008zx} for typical lifetimes). This renders the existence of such deeper vacua
effectively harmless and the dilaton domination scenario phenomenologically viable (see also \cite{Abel:2000bj}).

In this paper we reconsider the MSSM with dilaton dominated soft terms
and then extend the analysis to the NMSSM. In the MSSM there are only 
two independent parameters after  electroweak
symmetry breaking has been implemented. 
Within this two-dimensional parameter space one can obtain the 
observed Higgs mass for large $\tan\beta$
and simultaneously satisfy all other current experimental constraints.
Nevertheless, the thermal relic abundance is generically too large and the requirement of a dark matter relic abundance 
in accord with observation severely restricts
the allowed parameter space, implying an upper bound on the superpartner masses which makes it fully testable at the LHC-14. 
This also implies that the leading radiative corrections to the SM-like Higgs mass cannot be large enough to push the Higgs mass to
its observed central value, resulting in a Higgs mass  that is acceptable given theoretical and experimental uncertainties, but slightly low.

As mentioned above, singlet extensions benefit from a sizeable additional contribution to the diagonal component of the tree-level Higgs mass
matrix in the small $\tan\beta$ regime if the MSSM-singlet coupling $\lambda$ is large. This will result in an enhanced lightest Higgs mass
if the mixing within the Higgs sector is small. While this is not possible in the usual $\Z3$ symmetric NMSSM, we will consider very well
motivated generalised versions of the NMSSM where small mixings can be natural, resolving the tension between the Higgs mass
and a viable dark matter sector. The mass scale of the MSSM superpartners is however still very constrained, as for
large $\lambda$ the singlet states have to be rather heavy in order to avoid large mixings, which 
in turn implies a bino-like LSP whose relic abundance can only be kept small enough for not too heavy MSSM states.
A possibility to circumvent this conclusion is the decoupling regime of small $\lambda$, where the LSP can be singlino-like
with the correct relic abundance while all MSSM states can be heavy.

This paper is organized as follows. In Section~\ref{sec:dilatonbc}
we introduce the MSSM and NMSSM and recall the  dilaton dominated soft terms.
In Section~\ref{sec:mssm} we compute the low energy spectrum for the MSSM
while in Section~\ref{sec:gnmssm} we repeat the analysis for the NMSSM.

\section{Dilaton dominated supersymmetry breaking}
\label{sec:dilatonbc}

In this section we review the high scale boundary conditions
for dilaton dominated supersymmetry breaking in the MSSM and NMSSM in order to
set the stage for the following sections.
Let us start with the MSSM.

\subsection{Dilaton domination in the MSSM}
The superpotential of the MSSM is given by
\begin{align}
 \mathcal{W}_{\rm MSSM}=&\mu_{h}H_{u}H_{d}  +\sum_{\text{generations}}\big(y_{U}Q_{L}U_{R}H_{u} \nonumber \\
&+
y_{D}Q_{L}D_{R}H_{d}+y_{L}L_{L}E_{R}H_{d}\big)\ ,
\end{align}
where $Q_L$ are the quark doublets, $U_R, D_R$ are the quark singlets
and $H_{u,d}$ are the two Higgs multiplets. 
For the MSSM the dilaton dominated soft supersymmetry breaking terms 
are a universal scalar mass $m_0$ (which coincides with the gravitino mass 
$m_{3/2}$) and a $B_h$-parameter generated
at some high scale such as $M_{\rm GUT}$.
The (canonically normalised) 
gaugino masses $m_{1/2}$ and the $A$-terms are also universal
and at leading order related to $m_0$ by \cite{Kaplunovsky:1993rd}
\begin{equation}
m_{1/2} = \sqrt{3}\,m_0 \ , \qquad
 A_0=-\sqrt{3}\, m_{0}\ .
\label{dilatonBC}
\end{equation}
Thus before imposing electroweak symmetry breaking 
there are three independent parameters $\mu_h, m_0$ and $B_h$.
Requiring correct electroweak symmetry breaking and imposing $v\sim246\gev$
will effectively fix one of these parameters (up to a discrete choice). 
In fact it is common practice to use the vacuum equations to replace $\mu_h$ and $B_h$ by 
$\tan\beta\equiv v_u/v_d$ and sign($\mu_h$), leaving $m_0,\tan\beta$ and sign($\mu_h$) as free parameters.
For the more constrained case where $\mu_h$ is generated by the Giudice-Masiero
mechanism \cite{Giudice:1988yz}
the original parameter space is only two-dimensional 
as the additional constraint  $B_h=2\mu_h m_0$ holds \cite{Kaplunovsky:1993rd}.
This latter case necessarily has a light Higgs 
which has been ruled out already for some time  \cite{Barbieri:1993jk,Brignole:1993dj}. 
We will therefore concentrate on the general case with three independent 
parameters in the next section and indeed find that it is compatible 
with current experimental constraints.

Before we proceed, let us briefly discuss the corrections 
to the universal scalar mass and the relations given in Eq.~\eqref{dilatonBC}. 
First of all there is the anomaly mediated contribution
to the gaugino masses $\delta m_{1/2}=\tfrac{b}{16\pi^2} m_0$ 
where $b$ is the one-loop coefficient of the $\beta$-function \cite{Randall:1998uk,Giudice:1998xp}.
In addition, there are loop corrections to the dilaton K\"ahler potential 
which induce further corrections to  $m_0$ and \eqref{dilatonBC} \cite{Louis:1994ht,Casas:1996zi}. 
The size of the latter is difficult to estimate and also depends on the specific scenario
under consideration. In the following we will assume a setup in which these model-dependent corrections
are small and the model independent tree-level relations \eqref{dilatonBC} hold to good approximation.

\subsection{Dilaton domination in singlet extensions}
\label{sec:dilatonbc_nmssm}

The most general extension of the MSSM by a gauge singlet chiral superfield $S$ has a superpotential of the form
\begin{eqnarray}
 \mathcal{W}_{\rm GNMSSM} &=& \mathcal{W}_{\rm MSSM} + \lambda S H_u H_d + \xi S \nonumber \\
 &&+ \tfrac{1}{2} \mu_s S^2  + \tfrac{1}{3}\kappa S^3\ .
\label{eq:wgnmssm}
\end{eqnarray}
where $\lambda, \xi, \mu_{s}$ and $\kappa$ are conventionally chosen as real parameters. As the $\Z3$ symmetric version with 
$\mu_h=\mu_s=\xi=0$ is usually referred to as {\it{the}} NMSSM, \footnote{Dilaton domination in this $\Z3$ symmetric NMSSM was also considered in \cite{Kraniotis:1995gk}.} we will denote this more general case as the GNMSSM.

The $\mathcal{W}_{\rm GNMSSM}$ given in Eq.~(\ref{eq:wgnmssm}) seems to have a `double' $\mu$ problem, 
as the SM symmetries do not prevent arbitrarily high scales for the dimensionful mass terms. 
However these terms can be naturally of order the 
supersymmetry breaking scale if there is an underlying $\Z{4}^R$ or $\Z{8}^R$ symmetry \cite{Lee:2011dya}.
In such a case the superpotential is identical to the $\mathbbm{Z}_3$ symmetric NMSSM before supersymmetry breaking with $\mu_h=\mu_s=\xi=0$. 
However, after supersymmetry breaking these additional 
superpotential terms are generated with a size that is set by the 
supersymmetry breaking scale in the visible sector, i.e.\
the gravitino mass $m_{3/2}$. 
Not only is the fine tuning particularly promising in this setup, but also the severe tadpole~\cite{Abel:1996cr} and domain wall~\cite{Abel:1995wk} 
problems of the $\mathbbm{Z}_3$ symmetric NMSSM are avoided. 
For broken supersymmetry also the discrete $R$ symmetry is broken but 
the subgroup $\Z{2}^{R}$, corresponding to the usual matter parity, remains 
unbroken \cite{Lee:2011dya}. 
As a result the lightest supersymmetric particle (LSP) is stable and a candidate for dark matter.

The general soft supersymmetry breaking terms associated with the Higgs and singlet sectors are
\begin{align}
 & V_\text{soft} 
   =  m_s^2 |s|^2 + m_{h_u}^2 |h_u|^2+ m_{h_d}^2 |h_d|^2 + \big(B_h \, h_u h_d \nonumber \\
   & + \lambda A_\lambda s h_u h_d + \tfrac{1}{3}\kappa A_\kappa s^3 + \tfrac{1}{2} B_s s^2  + \xi_s s + h.c.\big) \;.
\label{soft}
\end{align}
As in the MSSM, the soft supersymmetry breaking terms are universal at leading order in the dilaton domination scenario.
All soft scalar masses including the soft mass $m^2_s$ are given by a universal $m_0^2$
while the gaugino masses and $A$-terms obey \eqref{dilatonBC} unchanged,
\begin{align}
 &m_{1/2}=\sqrt{3} m_0, \quad A_0 = A_\lambda = A_\kappa = - \sqrt{3} m_0,  \nonumber \\
 & m_{h_u}=m_{h_d}=m_s=m_0 \;.
\label{dilatonBCN}
\end{align}
For the $B$ terms we again have two choices.
They are either independent or, in the more constrained situation when
$\mu_h$ and $\mu_s$ are generated by the Giudice-Masiero mechanism, related
by $B_h = 2 \mu_h m_0, B_s=2 \mu_s m_0$. 
Thus before electroweak symmetry breaking is imposed the parameter space is either nine- 
or seven-dimensional with parameters $(m_0,\mu_h, B_h,\lambda,\kappa,\mu_s,B_s,\xi,\xi_s)$
or $(m_0,\mu_h,\lambda,\kappa,\mu_s,\xi,\xi_s)$ respectively. Requiring 
the correct electroweak symmetry breaking and imposing $v\sim246\gev$
will effectively fix one of these parameters. In addition the vacuum equations 
allow us  to replace two of the input parameters of the model by 
$\tan\beta$ and the vacuum expectation value (VEV) of $s$ which we denote as
$v_s$. 
Moreover, for the case that all variables are taken to be independent, i.e.\ in the case where no constraint on the $B$-terms is applied,
one of the dimensionful parameters can be eliminated by a shift in $v_s$, which we take to be $\xi$, i.e.\ we choose $\xi=0$. 
In summary, after electroweak symmetry breaking one ends up with seven independent input parameters in the general case, e.g.\ $(m_0,\tan\beta,\lambda,\kappa,\mu_s,B_s,v_s)$
and six in the constrained situation, $(m_0,\tan\beta,\mu_h,\lambda,\kappa,\mu_s)$.

\section{Phenomenology}
\subsection{SUSY, Higgs and DM cuts}
\label{sec:cuts}

In this section we briefly describe the cuts that we impose for the numerical analyses of the MSSM and the GNMSSM.
For the (first generation of) squarks and gluinos we use a cut of $m_{\tilde{q}}>1.7\tev$ and $m_{\tilde{g}}>1.7\tev$.
We further require the chargino and slepton masses to be above $100\gev$.
We also require that the lightest supersymmetric particle (LSP) is a neutralino which is a good dark matter candidate and its
relic density is within the $5\sigma$ PLANCK \cite{Ade:2013zuv} range of $0.1064 \le \Omega h^2 \le 0.1334$.
In addition to constraints from the relic density there are constraints from dark matter direct detection searches which limit the cross section of the lightest neutralino
to nucleons.
We require that the points are consistent with dark matter direct detection constraints, in particular with the latest constraint from LUX \cite{Akerib:2013tjd}. 
Finally, for the Higgs mass we take the average of the CMS and ATLAS best fit values of $125.7 \gev$ \cite{CMS-PAS-HIG-13-005} and $125.5 \gev$ \cite{ATLAS-CONF-2013-014} respectively and allow for a $3\gev$ uncertainty, $m_h = 125.6 \pm 3\gev$.

For our numerical analyses we use \SPheno \cite{Porod:2003um,Porod:2011nf} created by \SARAH \cite{Staub:2008uz,Staub:2009bi,Staub:2010jh,Staub:2012pb}.  This version performs a complete one-loop calculation of all SUSY and Higgs masses \cite{Pierce:1996zz,Staub:2010ty} and includes the dominant two-loop corrections for the scalar Higgs masses \cite{Dedes:2003km,Dedes:2002dy,Brignole:2002bz,Brignole:2001jy}. The dark matter relic density as well as the direct detection bounds are calculated with \MO \cite{Belanger:2006is,Belanger:2007zz,Belanger:2010pz}.

\subsection{The MSSM} 
 \label{sec:mssm}
\begin{figure}[!h!]
 \centering
 \includegraphics[width=0.69\linewidth]{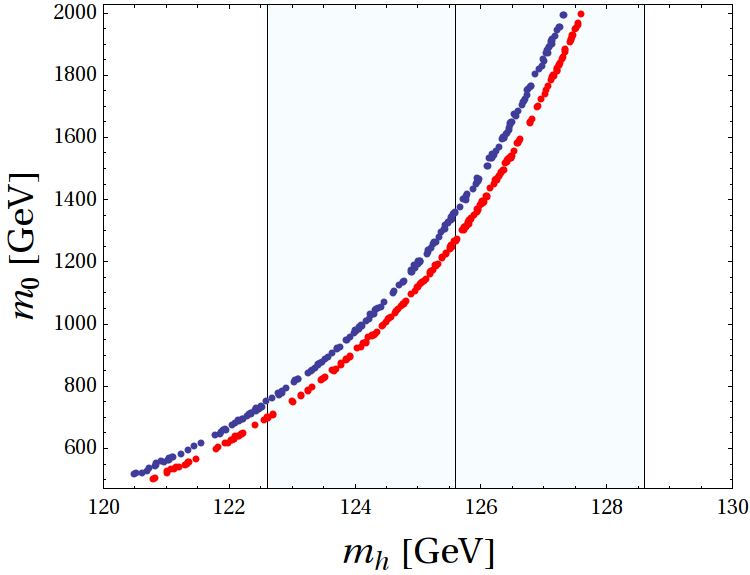}  
 \caption{Higgs mass as a function of $m_0$ for large $\tan\beta$ and $\mu_h >0$ (red) and $\mu_h <0$ (blue) in the MSSM.
 (For interpretation of the references to colour in this figure legend, the reader is referred to the web version of this article.)}
 \label{fig:1}
 \end{figure}
\noindent
\begin{figure}[!t!]
 \includegraphics[width=0.49\linewidth]{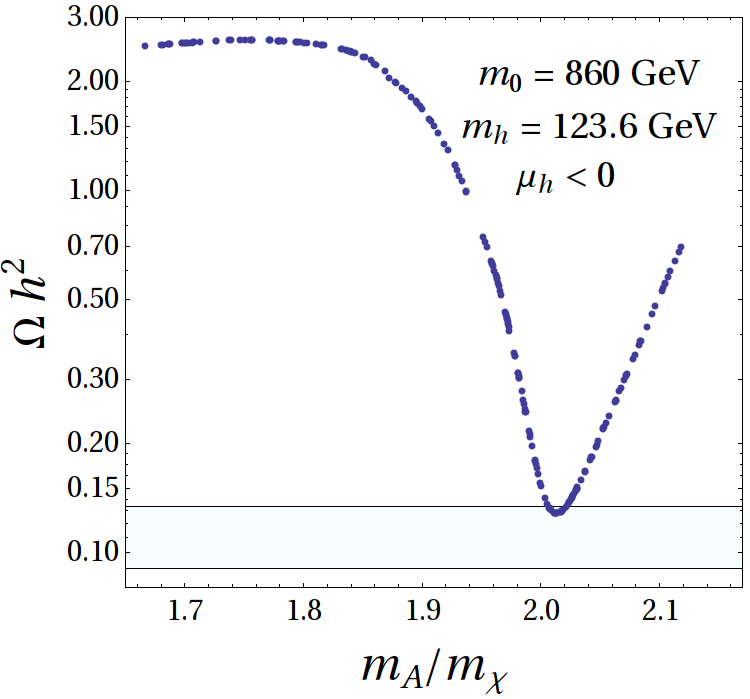}  
 \includegraphics[width=0.49\linewidth]{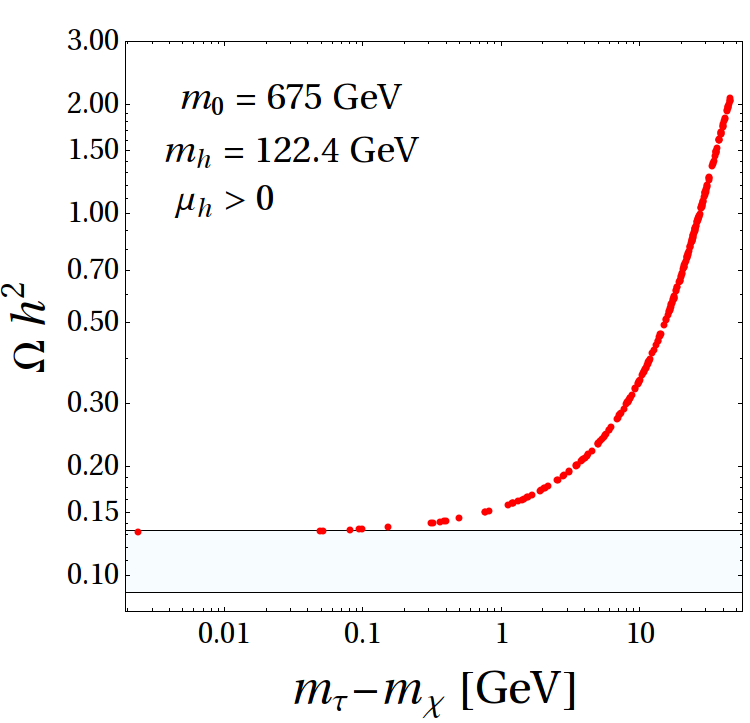} 
 \caption{The MSSM case. Left panel: $\mu <0$. Right panel: $\mu >0$.
  A small enough relic abundance can only be achieved for sufficiently small values of $m_0$ -- we show the critical values for both cases.
  It can be seen that in order to achieve a sufficiently small relic abundance the process has to be resonant with $m_A \sim 2 m_\chi$ for $\mu_h <0$ 
  or in the stau coannihilation region in the case of $\mu_h >0$. Only in the case $\mu_h <0$ do we achieve a Higgs mass consistent within the uncertainties.}
 \label{fig:2}
 \end{figure}

\begin{figure}[!t!]
 \includegraphics[width=0.69\linewidth]{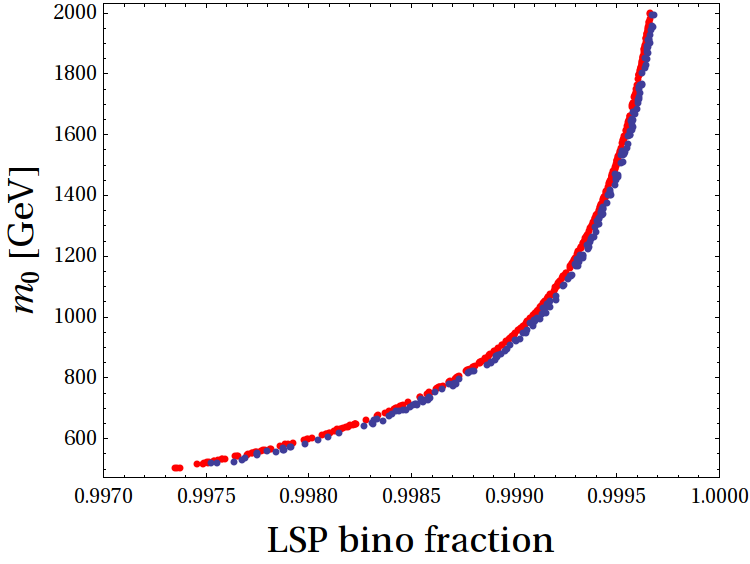} 
 \caption{LSP bino fraction as a function of $m_0$ for large $\tan\beta$ and $\mu_h >0$ (red) and $\mu_h <0$ (blue) in the MSSM.
}
 \label{fig:bino}
 \end{figure}
As explained in Sec.~\ref{sec:dilatonbc} our independent input parameters are $m_0, \tan\beta$ and sign$(\mu_h)$.
In Fig.~\ref{fig:1} we show the resulting dependence of the Higgs mass on $m_0$ for the case that $\tan \beta$ is large.
It can be seen that there is no problem to achieve a large enough Higgs mass via radiative corrections.\footnote{This of course ignores the fine tuning problem.}
In fact for $\tan\beta > 10$ there is almost a one-to-one correspondence between $m_0$ and the resulting Higgs mass for a given sign($\mu_h$). 
The minimal values of $m_0$ to achieve a Higgs mass
consistent with observation within the theoretical and experimental uncertainties are $m_0=695 \gev$  ($m_0=760 \gev$) for $\mu_h>0$ ($\mu_h<0$), well above the value of
$m_0 \gtrsim 500 \gev$ required by the SUSY cuts.
The precise numbers are however rather sensitive to the top Yukawa coupling, which enters the dominant radiative corrections.
If we were to take the one sigma upper limit on the top mass rather than its central value, 
radiative corrections are somewhat increased and the lower bound on $m_0$ is relaxed to $m_0=615 \gev$  ($m_0=660 \gev$) for $\mu_h>0$ ($\mu_h<0$).
In the following discussion we will stick to the central value of the top mass.

Turning to the neutralino sector, we find that the LSP is always an almost pure bino, with mass $m_{\tilde{\chi}_0^1} \simeq 0.8 \cdot m_{3/2}$,
somewhat below the gravitino mass.
The fact that the gravitino is rather close in mass to the neutralino is potentially dangerous cosmologically.
Whether this scenario is viable or not however depends on the cosmological history of our Universe, which we are agnostic about in this 
letter. Independently, a bino LSP implies that the thermal relic abundance is generically too large because
of a rather small annihilation cross section. One therefore has to go to special regions in parameter space where the relic
abundance can be small. 
Generically there is a tension between large $m_0$ implying a heavy SUSY spectrum (needed in this setup to achieve the correct Higgs mass) and a small enough
relic abundance.
One possible option would be to have coannihilations with another particle such as the stau, if the mass difference between neutralino and stau
is sufficiently small. This can be achieved even for our very restrictive boundary conditions (for positive $\mu_h$), but just about falls short of raising 
the Higgs mass into the observed window. In fact for the case of positive $\mu_h$ the maximal value for $m_0$ for which the correct relic abundance can be achieved is
$m_0 \simeq 675 \gev$, whereas the minimal value to obtain a large enough Higgs mass is $m_0 \simeq 695 \gev$.
Another option to deplete the relic abundance would be to have a mediating particle close to resonance, e.g.\ $m_A \sim 2 m_\chi$.
This can also be realised (for negative $\mu_h$) and turns out to be more promising.
In this case we do find viable points within the theoretical uncertainties, with an allowed range of $760 \gev \le m_0 \le 875 \gev$.

Given that the requirement of a relic abundance in accordance with observation is crucial to establishing an upper bound on the superpartner mass
scale, let us discuss in some more detail how this comes about. The annihilation cross section which sets the relic abundance is maximised
on the pole, where it is proportional to $g_{\chi\chi A}^2/\Gamma_A^2$, with $g_{\chi\chi A}$ the LSP-LSP-Higgs coupling and $\Gamma_A$ the width of the mediating Higgs scalar.
The width $\Gamma_A$ scales directly with $m_A$ and hence with $m_0$, decreasing the annihilation cross section with increasing SUSY breaking scale.
Furthermore the coupling $g_{\chi\chi A}$ originates from the Higgs-Higgsino-Gaugino term and is zero for a pure bino.\footnote{One might worry that in the case of an
almost pure bino where the tree-level coupling is very suppressed the one-loop coupling (see e.g.~\cite{Djouadi:2001kba}) might give a non-negligible correction. We find however 
that for the case of interest the one-loop coupling gives a correction of $\mathcal{O}(\%)$ and hence should not change the picture appreciably.}
The coupling is therefore proportional to the (small) Higgsino component of the LSP, which decreases with increasing $m_0$ (see Fig.~\ref{fig:bino}).
The combination of these two effects quickly makes sufficiently efficient annihilation impossible, even on the pole, resulting in this rather stringent upper bound on $m_0$. 

This leaves us with a very predictive scenario for dilaton domination in the MSSM, as all superpartner masses scale directly with $m_0$.
In particular the allowed range of $m_0$ corresponds to gluino and squark masses of $ 2750 \gev \lesssim m_{\tilde{g}}  \lesssim 3150 \gev $ and  $ 2450 \gev \lesssim m_{\tilde{q}} \lesssim 2800 \gev$ 
respectively, which should be fully testable with the upcoming LHC-14 run \cite{ATLAS:2013hta,Baer:2013fva}. While this is a viable scenario given the theoretical and experimental uncertainties,
an obvious shortcoming is that it is not possible to raise the Higgs mass to its observed central value.
This is different in singlet extensions, which we will discuss next.

\subsection{Singlet extensions}
\label{sec:gnmssm}

\begin{figure*}[!h!]
\begin{center}
 \includegraphics[width=0.29\linewidth]{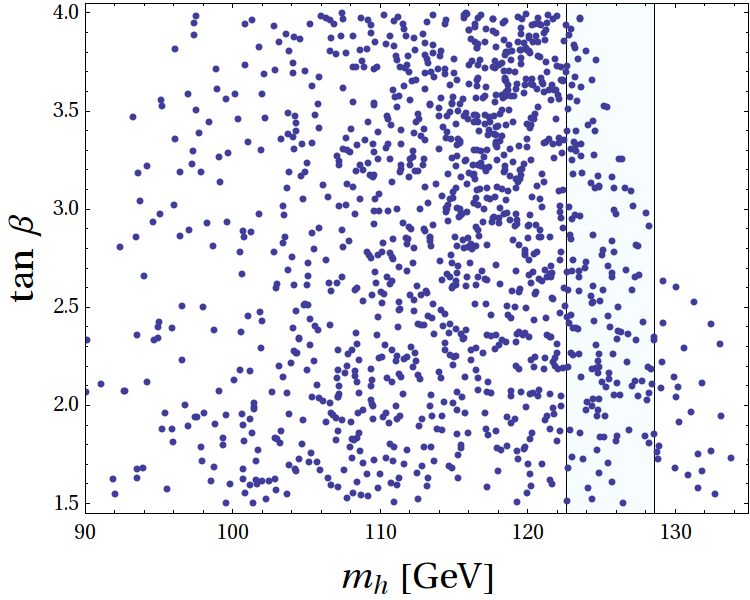}   
 \includegraphics[width=0.29\linewidth]{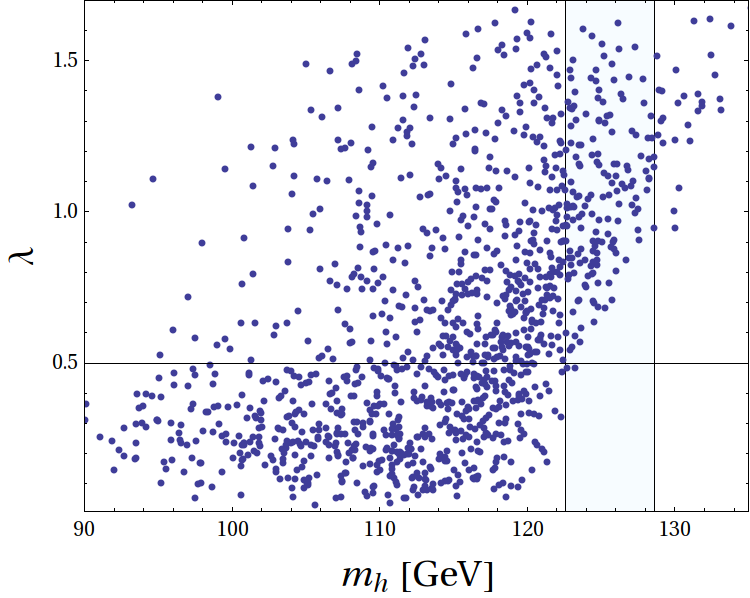} 
 \includegraphics[width=0.3\linewidth]{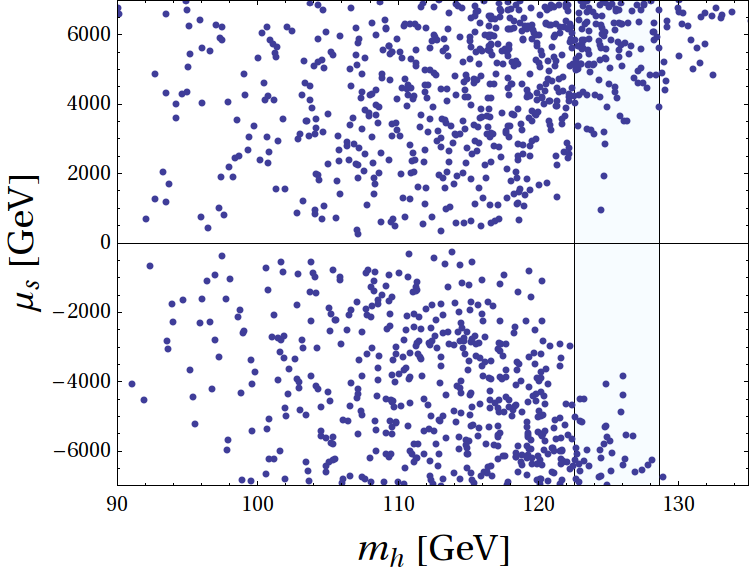}
\end{center}
 \caption{Dependence of the standard model like Higgs mass on $\tan\beta$, $\lambda$ and $\mu_s$ for the GNMSSM. For these figures we have applied the
 appropriate SUSY cuts, but not the cut on the dark matter relic abundance.}
 \label{fig:gnmssm}
 \end{figure*}
\begin{table*}[t]
\centering
 \begin{tabular}{|l|c c c c c|}
\hline
 & MSSM & GNMSSM1 & GNMSSM2 & GNMSSM3 & GNMSSM4\\
\hline
$m_0$ [GeV]                                 & 860                    & 500                   & 500                   & 500                   &1400\\
$\tan \beta$                                & 36.3                   & 2.7                   & 2.8                   & 2.7                   &33\\ 
$\mu_h$  [GeV]                              & tad                    & 1060*                 & 1000                  &1350*                  &2700*\\
$B_h~[\text{GeV}^2]$                        & tad                    & $2.7\cdot10^5$*       & $2\, m_0 \, \mu$      &$9.7\cdot10^5$*        &$2.6\cdot10^5$*\\
$\lambda$                                   & --                     & 1.5                   & 1.5                   & 1.5                   &0.003\\
$\kappa$                                    & --                     & -0.5                  &-0.58                  &-0.5                   &0.14\\
$v_s$       [GeV]                           & --                     & 500                   & 1330*                 & -102*                 &3440\\
$\mu_s$     [GeV]                           & --                     & -5000                 & 5400                  &-5000                  &-128\\   
$B_s~[\text{GeV}^2]$                        & --                     & $2850^2$              &$2\, m_0 \, \mu_s$     &$8.2\cdot 10^6$        &$6.3\cdot 10^5$\\
$\xi~[\text{GeV}^2]$                        & --                     & 0                     & $-2.5\cdot10^6$*      &0                      &0\\
$\xi_s~[\text{GeV}^3]$                      & --                     & $-3.7\cdot 10^9$*     & $-3.3\cdot 10^9$*     &0                      &$-4.3\cdot 10^9$*\\
\hline  
$m_{\tilde{g}}$  [GeV]                      &3100                    &1900                   & 1900                  &1900                   &4800\\      
$m_\text{squark}$ [GeV]                     &2100-2900               &1600-1800              & 1300-1750             &1300-1750              &3300-4500\\     
$m_\text{slepton}$ [GeV]                    &800-1300                &600-750                & 600-750               &600-750                &1300-2050\\     
$m_{\tilde{\chi}^\pm_1}$ [GeV]              &1250                    &710                    & 710                   &710                    &2050\\    
\hline                         
$m_{h_1}$ [GeV]                             &123.4                   &125.7                  & 123.6                 &127.2                  &125.9\\   
$m_{h_2}$ [GeV]                             &1385                    &1600                   & 1550                  &1580                   &1140\\   
$m_{A_1}$ [GeV]                             &1410                    &785                    & 785                   &785                    &1680\\    
\hline
$m_{\tilde{\chi}^0_1}$ [GeV]                &700                     &390                    &390                    &390                    &540\\    
$\tilde{\chi}^0_1$ bino part                & 0.999                  &0.998                  & 0.998                 &0.998                  &$\mathcal{O}(10^{-9})$\\
$\tilde{\chi}^0_1$ wino part                & $\mathcal{O}(10^{-6})$ &$\mathcal{O}(10^{-4})$ &$\mathcal{O}(10^{-4})$ &$\mathcal{O}(10^{-4})$ &$\mathcal{O}(10^{-10})$\\
$\tilde{\chi}^0_1$ higgsino part            & 0.001                  &0.002                  & 0.002                 &0.002                  &$\mathcal{O}(10^{-8})$\\
$\tilde{\chi}^0_1$ singlino part            & --                     &$\mathcal{O}(10^{-5})$ &$\mathcal{O}(10^{-5})$ &$\mathcal{O}(10^{-5})$ &1\\
$\Omega h^2$                                &  0.13                  &0.11                   &0.12                   &0.11                   &0.11\\
$\sigma_p [\text{cm}^2]$                    & $ \sim 10^{-48}$       & $\sim 10^{-46}$       &$\sim 10^{-46}$        &$\sim 10^{-46}$        &$\sim 10^{-49}$\\
\hline
 \end{tabular}
\caption{Benchmark points for the MSSM and the GNMSSM. All input parameters except $\tan\beta$ and $v_s$ are given at the GUT scale. Values marked with a * are output values at the electroweak scale determined by the electroweak symmetry breaking conditions.}
\label{tab:benchmark}
\end{table*}

The superpotential and soft terms of the general NMSSM have been given in Sec.~\ref{sec:dilatonbc}.
Before going to the most general case, let us briefly discuss some restricted cases that might be of interest. 
To start with, the very well known  $\mathbbm{Z}_3$ symmetric NMSSM requires $A_\kappa^2 \gtrsim 9 m^2_s$ (see e.g.\ \cite{Ellwanger:2009dp})
in order to have a real vev for $S$, which is not possible in the dilaton domination scenario as the relations in \eqref{dilatonBCN} show. 
Another problem is that the region of large $\lambda$ and small $\tan\beta$ which is preferred due to fine tuning considerations is
phenomenologically unacceptable as mixing effects decrease the lightest Higgs mass, overcompensating the additional tree-level contribution.
That the $\mathbbm{Z}_3$ symmetric NMSSM does not work in the context of dilaton domination was also found in \cite{Kraniotis:1995gk}.
In contrast the GNMSSM allows for a natural limit of heavy singlet states, suppressing the mixing in the Higgs sector.
In Tab.\ref{tab:benchmark} we show example points for some cases of interest, including a case in which both linear terms are set to zero, $\xi=\xi_s=0$, and where the constraints on the B-terms hold, $B_h = 2 \mu_h m_0, B_s=2 \mu_s m_0$. 

To illustrate the general dependence on the different input parameters we performed a scan for the general case 
in the region of interest corresponding to small $\tan\beta$ and large $\lambda$.
In Fig.~\ref{fig:gnmssm} we show the dependence of the Standard Model like Higgs mass on the input parameters $\tan\beta$, $\lambda$ and $\mu_s$,
which are most relevant. Note that $\lambda$ and $\mu_s$ are the GUT scale parameters. One can clearly see the increased Higgs mass for small $\tan\beta$ and large $\lambda$.
It can also be seen that the singlet mass term $\mu_s$ has to be sufficiently large to avoid too large mixing in the Higgs sector.

For the neutralino sector this implies that the LSP is again a bino-like neutralino, as the singlino is typically rather heavy.
To achieve the correct relic abundance one is therefore forced to special regions as in the MSSM case.
Given that the stau coannihilation region cannot be accessed for small values of $\tan\beta$ with the given boundary conditions, 
the only remaining option is resonant annihilation via the Higgs funnel, implying $m_A\sim 2 m_\chi$. 
It turns out that for the GNMSSM, due to the additional tree-level contribution, there is no tension between a large enough Higgs mass and the relic abundance. 
Nevertheless, the mass scale of the MSSM superpartners is still very constrained, as 
the annihilation channel ceases to be effective for too large $m_0$, as the width as well as the bino fraction increases, similar to the MSSM 
case.\footnote{For an explicit expression of the LSP-LSP-Higgs coupling in the case of general MSSM singlet extensions (which trivially reduces to the MSSM case) 
see e.g.\ the appendix of \cite{SchmidtHoberg:2012ip}.}
We find it remarkable that in spite of the many free parameters, the dilaton domination turns out to be rather predictive even in general singlet extensions 
of the MSSM in the large $\lambda$ regime.
The only possibility to circumvent this conclusion is to go to the decoupling regime of small $\lambda$, where the LSP can be singlino-like
with the correct relic abundance while all MSSM states can be heavy. We also checked the direct detection cross section for all cases and find
that they are significantly below the current bound set by LUX, which is not surprising given that the predominant LSP component is either
bino or singlino.

\section*{Acknowledgements}
We would like to thank Florian Staub for very useful discussions. We would also like to thank Ricardo Marquez for his persistent help and support.
J.L.\ thanks the Weizmann Institute for their kind hospitality during the
final stages of this work.
This work was supported by the German Science Foundation (DFG) within the
Collaborative Research Center (SFB) 676 
``Particles, Strings and the Early Universe'', G.I.F.\ --
the German-Israeli Foundation for Scientific Research and Development
and the I-CORE program of the Planning and Budgeting
Committee and the Israel Science
Foundation (grant number 1937/12).

\bibliography{NMSSM}
\bibliographystyle{ArXiv}

\end{document}